\documentclass[a4paper,12pt]{myarticle}
\usepackage{graphicx}
\renewcommand\figurename{\small {\sc Fig.}}
\renewcommand\tablename{\small {\sc Table}}
\begin{document}
\thispagestyle{empty}
\ \vspace{0cm}
\begin{center}
\renewcommand{\baselinestretch}{1.1} \normalsize
{\bf \large 
\hbox{\hspace{-1cm}Exponential Weighting and Random-Matrix-Theory-Based Filtering}
of Financial Covariance Matrices for Portfolio Optimization} \par
\vspace{0.7cm}
{Szil\'ard Pafka$^{1,2,a}$, Marc Potters$^{3,b}$ and Imre Kondor$^{1,4,c}$} \par \vspace{0.35cm}
{\it \small $^1$Department of Physics of Complex Systems, E\"otv\"os
University \\
P\'azm\'any P.\ s\'et\'any 1/a, 1117 Budapest, Hungary} \par
\vspace{0.1cm}
{\it \small $^2$Risk Management Department, CIB Bank \\
Medve u.\ 4--14., 1027 Budapest, Hungary} \par
\vspace{0.1cm}
{\it \small $^3$Science \& Finance, Capital Fund Management \\
6 boulevard Haussmann, 75009 Paris, France} \par
\vspace{0.1cm}
{\it \small $^4$Collegium Budapest, Institute for Advanced Study  \\
Szenth\'aroms\'ag u. 2., 1014 Budapest, Hungary} \par \vspace{0.45cm}
January, 2004 \par \vspace{1cm}
{\bf Abstract} \par \vspace{0.7cm}
\parbox{13cm}{\small
We introduce a covariance matrix estimator that both takes into account 
the heteroskedasticity of financial returns (by using an exponentially 
weighted moving average) and reduces the effective dimensionality of the
estimation (and hence measurement noise) via techniques borrowed from
random matrix theory. We calculate the spectrum of large exponentially
weighted random matrices (whose upper band edge
needs to be known for the implementation of the estimation) analytically, 
by a procedure analogous to that used for standard random matrices.
Finally, we illustrate, on empirical data, the superiority of the newly introduced estimator
in a portfolio optimization context over both the method of exponentially 
weighted moving averages and the uniformly-weighted 
random-matrix-theory-based filtering.
\par \vspace{0.3cm} {\it PACS:} 87.23.Ge; 05.45.Tp; 05.40.--a
\par \vspace{0.3cm} {\it Keywords:} financial covariance matrices,
exponentially weighted moving averages, estimation noise, random matrix theory, noise filtering, 
portfolio optimization, risk management
} \end{center} \vspace{1cm} \par
\rule{5cm}{0.4pt} \par
{\small {\it E-mail:} $^a$syl@complex.elte.hu,
$^b$marc.potters@science-finance.fr, $^c$kondor@colbud.hu
\newpage \setcounter{page}{1}
\renewcommand{\baselinestretch}{1.1} \normalsize


\section{Introduction}

Covariance matrices of financial returns play a crucial role in 
financial theory and also in many practical applications.
In particular, financial covariance matrices are the key input
parameters to Markowitz's classical portfolio selection problem 
\cite{markowitz} which forms the basis of modern investment theory.
For any practical use of the theory, it would therefore be necessary 
to obtain reliable estimates for the covariance matrices of real-life financial
returns (based on historical data). It was clear
from the very outset that the estimation problem of such matrices suffers from
the "curse of dimensions": if one denotes by $N$ the number of
assets and by $T$ the length of the time series used for estimation,
one has to estimate $O(N^2)$ parameters from a sample of $O(NT)$ historical 
returns, and usually the condition $T\gg N$ cannot be fulfilled in
realistic financial applications. For finite $N$ and $T$,
with $N$ large and $T$ bounded for practical reasons\footnote{Typically 
one wants to consider several hundreds of assets and has available daily 
financial data over a period of a couple of years at most.},
the estimation error of the covariance matrix can become so overwhelming
that the whole applicability of the theory becomes questionable.

This difficulty has been well known for a long time, see e.g.\ Ref.\ 
\cite{eltongruber} and the numerous references therein. The effect of 
estimation noise (in the covariance matrix of financial
returns) on the solution of the classical portfolio selection problem has 
been extensively studied, see e.g. Ref.\ \cite{econnoisedet}.
The general approach to reducing this estimation noise has been to impose some
structure on the covariance matrix in order to decrease the effective number of 
parameters to be estimated.
This can be done by using several methods. For example, various "models" have been 
introduced on "economic" grounds, such as single and multi-index models, 
grouping by industry sectors or macroeconomic factor models 
(see e.g.\ the numerous references in \cite{eltongruber}). 
Alternatively, "purely statistical" covariance estimation methods have been 
used too, such as principal component analysis or Bayesian shrinkage estimation 
\cite{bayes}. Several studies compare the performance of (some of)
these covariance estimation procedures (in the framework of classical portfolio
optimization problem), see e.g.\ Ref. \cite{estimcomp}.
The general conclusion of all these studies is that reducing the 
dimensionality of the problem by imposing some structure on the
covariance matrix may be of great help in reducing the effect of measurement
noise in the optimization of portfolios.

The problem of noise in financial covariance matrices has been put
in a new light by the findings of Ref.\ \cite{bouchaud} and the
following Refs.\ \cite{stanley,bouchaud2,stanley2}, obtained by
the application of random matrix theory.
These studies have shown that correlation matrices determined from
financial return series contain such a high amount of noise that,
appart from a few large eigenvalues and the corresponding eigenvectors,
their structure can be regarded as random (in the example analyzed in Ref.\
\cite{bouchaud} 94\% of the spectrum could be fitted by that of a purely 
random matrix). 
The results of Refs.\ \cite{bouchaud,stanley} not only showed 
that the amount of estimation noise in financial correlation
matrices is large, but also provided the basis for a technique that can
be used for an improved estimation of such correlation matrices.
A "filtering" procedure based on "eliminating" those eigenvalues and eigenvectors
of the empirical correlation matrix that correspond to the noise band deduced 
from random matrix theory has been introduced in Refs.\ \cite{bouchaud2,stanley2}
and found to be very effective in reducing the estimation noise in
the portfolio optimization context.

In this paper we introduce a covariance matrix estimator that combines 
the filtering procedure based on random matrix theory (that seeks to 
attenuate estimation noise by reducing the effective dimensionality 
of the covariance matrix)
with the technique of exponentially weighted moving averages of returns
(that tries to take into account the heteroskedasticity
of volatility and correlations, a salient feature of real-life financial returns).
We show that this estimator can be very powerful in constructing portfolios with
better risk characteristics.
In particular, it seems that by taking into account 
the non-stationarity of the time-series, the estimator can outperform the standard
random matrix theory-based filter (where returns from a given time window are 
uniformly weighted).
Most remarkably, the spectrum of exponentially weighted random 
matrices (whose upper band edge needs to be computed for the
practical implementation of the estimation procedure) can be computed
analitically in a certain limiting case.
Concerning another aspect of the use of empirical financial covariance
matrices, we argue that even though matrices
obtained by simple exponential weigthing (without any noise filtering)
can be successful for determining the risk of a given portfolio, 
their use in the context of portfolio optimization can be very 
dangerous.


\section{A Covariance Matrix Estimator Based on Exponential Weighting and
Random Matrix Theory}

\newcommand{\subs}[1]{{\mbox{\scriptsize #1}}}
\newcommand \rmi {\mbox{i}}
\newcommand \rmd {\mbox{d}}
\newcommand \be {\begin{equation}}
\newcommand \ee {\end{equation}}
\newcommand \ba {\begin{eqnarray}}
\newcommand \ea {\end{eqnarray}}

Suppose we have a sample of (say, daily) returns of $N$ financial assets over a given period
of time. Let us denote by $x_{ik}$ the return of stock $i$ ($i=1,2,\ldots,N$) 
at time $t-k$, with $t$ the last point of time in the available data ($k=0,1,\ldots$). 
A simple and widely used estimator for the covariance matrix of returns is
the "sample" or "historical" matrix:
\be
C_{ij}=\frac{1}{T} \sum_{k=0}^{T-1} x_{ik} x_{jk},
\ee
where $T$ is the length of the sample. Under the assumption that the distribution of returns is 
Gaussian, this is also
the maximum likelihood estimator which is known to perform well 
in the limit $N=\mbox{fixed}$, $T\to\infty$.
However, in finite length samples, especially when $N$ is large,
the estimation noise (measurement error) can become significant. 
Estimation methods that can reduce this measurement error
have long been in the focus of attention of academics and 
practitioners alike. The root of the problem is evident: $N$ time series of length $T$ each
do not contain sufficient information to allow the $O(N^2)$ matrix 
elements to be reliably estimated unless $T\gg N$, which hardly ever occurs in a banking context.
Since the length of the time series is, for obvious reasons, rather limited 
(of the order of a few years, i.e.\ in cases with daily data 
$N\sim 1000$ at most), the only conceivable solution is 
to reduce the effective number of dimensions of the problem.
Over the years several techniques have been introduced and succesfully applied that
reduced the estimation error through a shrinking of dimensions. One of the latest 
of these techniques was inspired by results from random matrix theory.
It consists of "cleaning" the covariance matrix by
retaining only those components (eigenvalues and corresponding eigenvectors)
that are outside the noise band that corresponds to the spectrum of a purely 
random matrix (see Refs.\ \cite{bouchaud2,stanley2}).
It has been demonstrated empirically in Refs.\ \cite{bouchaud2,stanley2}
and subsequently via simulations in Ref.\ \cite{noisy23} that this technique
is indeed very powerful for reducing the estimation noise of covariance matrices used in
standard (mean--variance) portfolio optimization.

However, it is well known that financial returns exhibit heteroskedastic volatility
and correlations (i.e.\ the random processes generating the returns are not stationary). 
Accordingly, a large part of the 
financial academic literature has focused on modelling the dynamics of 
the covariance of financial returns\footnote{The most widely used approach to
modelling the heteroskedasticity of financial returns is via ARCH/GARCH processes, see e.g.\
Ref.\ \cite{arch} for a review.}. 
These ideas have also found their way into industrial practice:
in the early 1990's J.P.\ Morgan and Reuters introduced RiskMetrics 
\cite{riskmetrics}, a methodology for the determination of the market risk of portfolios.
RiskMetrics has soon become the most widely used method for measuring market
risk and it is now considered a benchmark in risk management. At the heart of the 
method lies the estimation of the covariance matrix of returns
("risk factors") through an exponentially weighted moving average\footnote{
The idea of the method is that old data become gradually obsolete, therefore they should
contribute less to the estimates than more recent information. The exponential weighting is 
consistent with GARCH, see Ref.\ \cite{riskmetrics}.}:
\be
\label{eq:ewrm}
C_{ij}=\frac{1-\alpha}{1-\alpha^T} \sum_{k=0}^{T-1} \alpha^k x_{ik} x_{jk},
\ee
where the normalization factor can be approximated
by $1-\alpha$ if $T$ is large enough.
This method has been found to be very successful in estimating the market
risk of given, fixed portfolios. 

In contrast, if this covariance matrix estimate is used
for portfolio optimization (i.e.\ for selecting the portfolio 
in a mean--variance framework, which involves the inversion of the matrix), 
the estimation error will be quite large for 
typical values of the ratio $T/N$ (see Ref.\ \cite{noisy23}). In the case of 
exponential weighting, the results in Ref.\ \cite{noisy23} imply that the 
degree of suboptimality will depend on the ratio of the effective time length
$-1/\log\alpha$ and the number of assets $N$. In particular, since the effective
time corresponding to the value of the exponential decay factor $\alpha$
suggested by Ref.\ \cite{riskmetrics} ($\alpha=0.94$ for daily data) is shorter than
the length of the time windows used in a typical standard (uniformly weighted)
covariance matrix estimation, it can be expected that for the same portfolio
size $N$ the effect of noise (suboptimality of optimized portfolios) will
be larger with exponential weighting than without it.
Nevertheless, dimension reduction techniques based on
random matrix theory (developed in Refs.\ \cite{bouchaud2,stanley2}
for uniformly weighted matrices) can be expected to be useful in reducing the 
effect of noise also in the case of exponential weighting.

The purpose of this paper is to introduce an estimator that, by the adaptation
of the filtering procedure of Refs.\ \cite{bouchaud2,stanley2} to
exponentially weighted matrices, can reduce the estimation noise of the
covariance matrix used for portfolio optimization (while it will still be able to 
take account of the non-stationarity of financial returns). The usefulness
of this procedure for portfolio optimization will also be illustrated.
In fact, since the spectrum of exponentially weighted 
purely random matrices of the form of Eq.\ (\ref{eq:ewrm}) (with $x_{ik}$ iid random variables)
will be seen to be
qualitatively similar to that of the standard (uniformly weighted) random
matrices, the same filtering procedure can be applied in both cases. 
The only difference lies in
the value of the upper edge of the noise spectrum. 
Therefore, in order to apply the filtering procedure, one has to know
the value of the upper edge of the noise
spectrum of an exponentially weighted purely random matrix for a given $N$ and $\alpha$. This can be
determined for each given $N$ and $\alpha$ by Monte Carlo simulation,
but most remarkably, in the limit of $N\to\infty,\alpha\to 1, 
N(1-\alpha)=\mbox{fixed}$ it is possible to obtain the full spectrum 
as the solution to a set of
analytical equations, as shown below.


\section{The Spectrum of Exponentially Weighted Random Matrices}

The derivation of the spectrum of exponentially weighted random matrices
follows the steps and notation of the standard (Wishart) 
case in Ref.\ \cite{marcref1} which is itself based on Ref.\ \cite{marcref2}.
In the limit of an infinite window, the exponentially weighted random matrix is
given by
\be
C_{ij}=\sum_{k=0}^{\infty}  (1-\alpha)\alpha^k x_{ik} x_{jk},
\ee
where $x_{ik}$ is assumed to be Gaussian iid with zero mean and
standard deviation $\sigma$.
One can rewrite 
\be
C_{ij}=\sum_{k=0}^{\infty} H_{ik} H_{jk} 
\ee
with $H_{ik}$ having a $k$-dependent variance $\sigma^2_k=\sigma^2(1-\alpha)\alpha^k$.

Following Ref.\ \cite{marcref1}, we can use the resolvent technique to write
the density of eigenvalues as the imaginary part of the derivative of a
log-partition function:
\be
\rho(\lambda)=\frac{1}{N\pi}\mbox{Im}\frac{\rmd}{\rmd \lambda}Z(\lambda),
\label{im_part}
\ee
with
\be
Z(\lambda)=-2\left.\log \int
\exp\left[-\frac{\lambda}{2}\sum_{i=1}^{N}\varphi_i^2
-\frac{1}{2}\sum_{i,j=1}^{N}\sum_{k=0}^{\infty}\varphi_i \varphi_j H_{ik}H_{jk}\right]
\prod_{i=1}^{N}\left(\frac{\rmd \varphi_i}{\sqrt{2\pi}}\right)\right. .
\ee

We can now average $Z(\lambda)$ over the random matrix $H_{ij}$. To keep the derivation
simple, we will average the argument of the log rather than average the log.
As in the standard case, one can use the formal device of the replica trick and show that
the result is indeed self-averaging. The $H_{ij}$-dependent term can be averaged using a
standard Gaussian integral:

\begin{eqnarray}
\left\langle\exp
\left[-\frac{1}{2}\sum_{i,j=1}^{N}\sum_{k=0}^{\infty}\varphi_i \varphi_j
H_{ik}H_{jk}\right]\right\rangle
&=&\prod_{k=0}^{\infty}\left(1-\sigma^2(1-\alpha)\alpha^k\sum_{i=1}^{N}\varphi_i^2\right)^{-1/2}\\
&=&\exp\left\{-\frac{1}{2}\sum_{k=0}^{\infty}
\log\left(1-\sigma^2(1-\alpha)\alpha^k\sum_{i=1}^{N}\varphi_i^2\right)\right\}.
\end{eqnarray}

We then introduce $q\equiv \sigma^2(1-\alpha)\sum
\varphi_i^2$  which we fix using an integral representation of the delta
function:
\be
\delta\left( q - \sigma^2(1-\alpha)\sum
\varphi_i^2 \right) 
=
 \int \frac{1}{2\pi} 
\exp\left[\rmi\zeta(q - \sigma^2(1-\alpha)\sum
\varphi_i^2)\right]
\rmd \zeta.
\ee
After performing the integral over the $\varphi_i$'s and writing $z=2\rmi\zeta(1-\alpha)$, we find:
\begin{eqnarray}
\nonumber
Z(\lambda)&=&-2\log \frac{NQ}{4\pi}
\int_{-\rmi\infty}^{\rmi\infty}
\int_{-\infty}^{\infty}
\exp\biggl[-\frac{N}{2}\biggl(\log(\lambda-\sigma^2z)+\\
&&\frac{1}{N}\sum_{k=0}^{\infty}\log(1-\alpha^kq)+
Qqz\biggr)\biggr]\,
\rmd q\,\rmd z,
\label{zl_final}
\end{eqnarray}
where $Q\equiv 1/(N(1-\alpha))$ measures the ``quality'' of the estimation as the ratio
of the decay time of the exponential weighting to the number of assets.

This is where the main difference with the Wishart case arises. The term $Q\log(1-q)$ is replaced
by
\be
F_Q(q)\equiv-\frac{1}{N}\sum_{k=0}^{\infty}\log(1-\alpha^kq),
\ee
which we need to compute in the  
$N\to\infty,\alpha\to 1$ limit with $Q\equiv 1/(N(1-\alpha))$ fixed.
We start by expanding the log in a Taylor series about 1
\begin{eqnarray}
F_Q(q)&=&\frac{1}{N}\sum_{k=0}^{\infty}\sum_{\ell=1}^{\infty} \frac{q^\ell\alpha^{k\ell}}{\ell}\\
&=&\frac{1}{N}\sum_{\ell=1}^{\infty} \frac{q^\ell}{\ell}\frac{1}{1-(1-Q/N)^\ell}.
\end{eqnarray}
Taking the $N\to\infty$ limit we find
\be
F_Q(q)=Q\sum_{\ell=1}^{\infty} \frac{q^\ell}{\ell^2}\equiv QF(q),
\ee
where $F(q)$ is the hypergeometric function with the property $F'(q)=-\log(1-q)/q$.

We can now perform the integrals over $z$ and $q$ using the saddle point
method, leading to the following equations:
\be
Qq=\frac{\sigma^2}{\lambda-\sigma^2z} \quad
\mbox{and}
\quad
 z=-\frac{\log(1-q)}{q}.\label{qandz}
\ee

Here we are saved by the fact that only the derivative ($-\log(1-q)/q$) of
the function $F(q)$ appears. 

To find the density we need to differentiate Eq.\ (\ref{zl_final}) 
with respect to $\lambda$.
Since we do not have explicit expressions for $q(\lambda)$ and
$z(\lambda)$ at the saddle point, it is important to realize that
partial derivatives with respect to these variables are zero.

We find:
\be
\frac{\rmd Z}{\rmd\lambda}
=
\frac{N}{\lambda-\sigma^2z(\lambda)}
=
\frac{NQq(\lambda)}{\sigma^2}.
\ee
We can now use Eq.\ (\ref{im_part}) to find the density of
eigenvalues:
\be
\rho(\lambda)=\frac{Q\mbox{Im}[q(\lambda)]}{\pi\sigma^2}.
\ee
Because Eqs.\ (\ref{qandz}) are transcendental, we cannot find an explicit form
for $\rho(\lambda)$, nevertheless it is straightforward to write $\rho(\lambda)$
as the zero of a single equation which can be solved numerically. We find
$\rho(\lambda)=Qv/\pi$ where $v$ is the solution of
\be
\frac{\lambda}{\sigma^2}-\frac{v\lambda}{\tan(v\lambda)}+\log(v\sigma^2)-\log\sin(v\lambda)-Q^{-1}=0.
\ee

The solution $\rho(\lambda)$ for a given $Q$ looks fairly similar to  $\rho(\lambda)$
of the standard (Wishart) case\footnote{For simplicity, in the following 
analysis we consider $\sigma=1$.}.
This is illustrated
in Fig.\ \ref{fig:spexpstd} where we plotted the spectrum of the exponentially weighted
random matrix with $Q\equiv 1/(N(1-\alpha))=2$ and the spectrum of the standard
random matrix with $Q\equiv T/N=3.45$ (for which the upper edges of the two
spectra coincide).
It can be clearly seen from the figure that the two curves run quite close to each other.

\begin{figure}[h!]
\begin{center}
\includegraphics[scale=0.65]{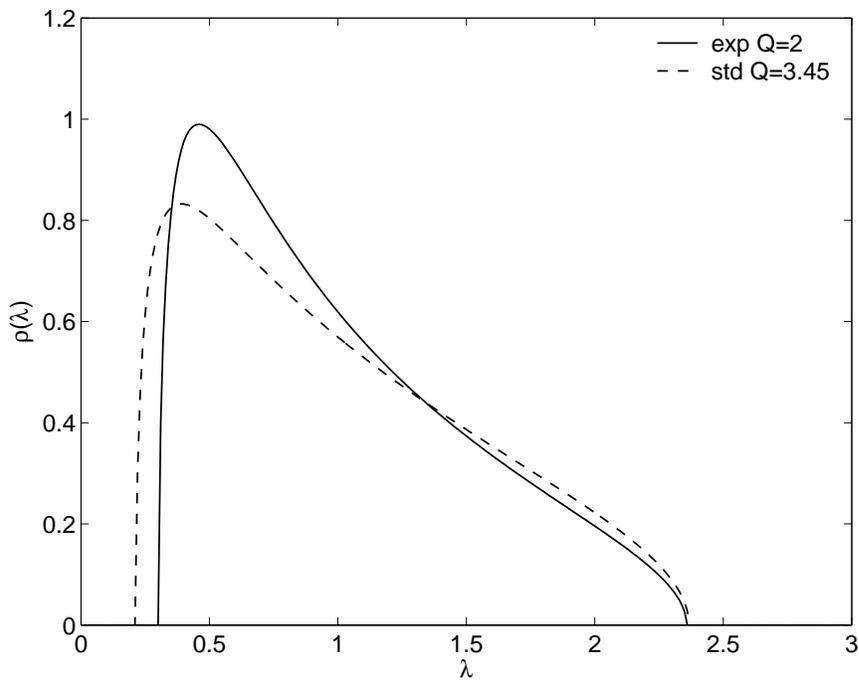}
\end{center}
\caption{Spectrum of the exponentially weighted
random matrix with $Q\equiv 1/(N(1-\alpha))=2$ and the spectrum of the standard
random matrix with $Q\equiv T/N=3.45$.
\label{fig:spexpstd}}
\end{figure}

The spectrum obtained in the limit of $N\to\infty$, $\alpha\to 1$ with 
$Q\equiv 1/(N(1-\alpha))$ fixed can be compared with the distribution
of eigenvalues for finite $N$. Fig.\ \ref{fig:spexpfinite} shows the
spectrum of the exponentially weighted random matrix with 
$Q\equiv 1/(N(1-\alpha))=2$ in the limit of $N\to\infty$ and the 
histogram of eigenvalues for one realization of the matrix with the
same value of $Q$ for finite $N=400$. It can be seen that the fit is quite good already for 
a single realization of the matrix.

\begin{figure}[h!]
\begin{center}
\includegraphics[scale=0.65]{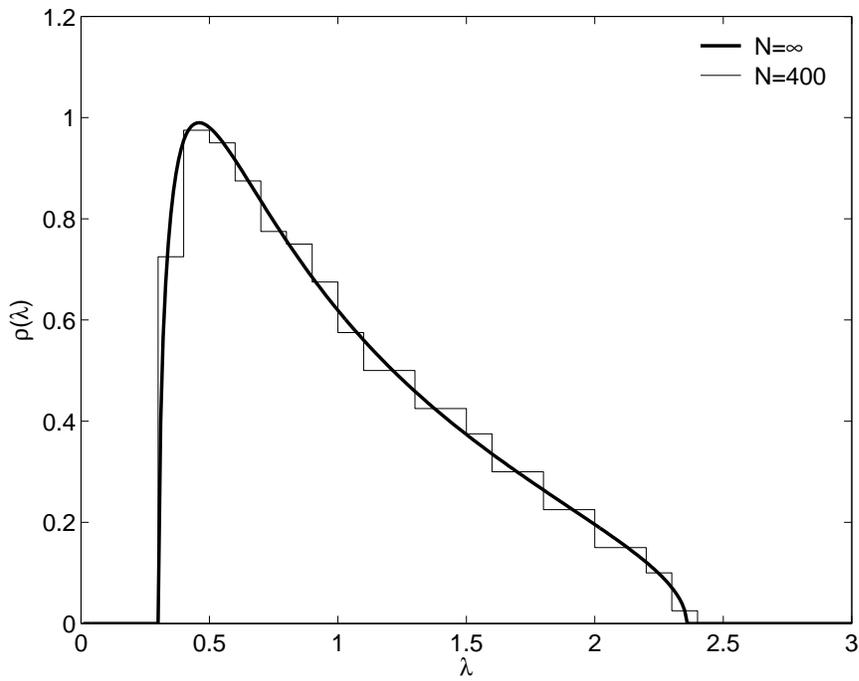}
\end{center}
\caption{Spectrum of the exponentially weighted
random matrix with $Q\equiv 1/(N(1-\alpha))=2$ in the limit 
$N\to\infty$ and the histogram of eigenvalues for one realization
of the matrix for finite $N=400$.
\label{fig:spexpfinite}}
\end{figure}


\section{Portfolio Optimization Results and Discussion}

In order to test the performance of the covariance matrix
estimator in the context of portfolio optimization, we consider
the simplest version of the classical (mean--variance) portfolio optimization 
problem: the portfolio variance 
$\sum_{i,j=1}^n w_i\,C_{ij}\,w_j$ is minimized under the
budget constraint $\sum_{i=1}^n w_i=1$, 
where $w_i$ denotes the weight of asset $i$ in the portfolio and
$C_{ij}$ the covariance matrix of asset returns.
In this case, the weights of the "optimal" (minimum variance) portfolio are simply
\be
\label{eq:optportf}
w_i^*=\frac{\sum_{j=1}^n C_{ij}^{-1}}{\sum_{j,k=1}^n C_{jk}
^{-1}}.
\ee
By eliminating all additional sources of uncertainty (such as, for example, 
expected returns that are notoriously hard to estimate) stemming from the
determination of different other parameters appearing in more complex 
formulations, this form provides an extremely convenient framework in which to test the efficiency
of different covariance matrix estimators as inputs for portfolio optimization (see 
Ref.\ \cite{noisy23}).

We assess the performance of several covariance matrix estimators
based on the out-of-the-sample performance of the portfolios constructed
using the covariance matrices provided by the estimators. 
For this purpose, we take a sample of financial returns (e.g.\ daily returns on stocks),
and we divide it into an estimation ("past") period and an evaluation 
("future") period.
We calculate different correlation matrix estimates based on returns only from 
the first period and we use them to construct "optimal" portfolios (as given
by Eq.\ (\ref{eq:optportf})).
Finally, we evaluate the performance of the estimators 
based on the standard deviation of the corresponding portfolio returns
in the second period.
In order to reduce the error that might arise from the use of a 
single sample, we perform our experiments on a large number of samples
bootstrapped from a larger dataset of daily stock returns. More precisely, starting 
from the same dataset of 1306 daily returns on 406 large-cap stocks traded on the 
New York Stock Exchange as used in Refs.\ \cite{bouchaud,bouchaud2},
for several values of $N$ (number of assets) and $T_2$ (the length of the 
evaluation period), in each iteration, 
we select at random $N$ assets and a period of time starting 
from the beginning of the dataset and ending at a random point in time 
(in the last third of the sample in order to have an estimation
period of sufficient length). The last $T_2$ data points of this bootstrapped
sample are used for evaluation, while the rest of the sample for estimation.

We consider several methods for estimating covariance matrices.
We calculate "historical" estimates based on uniformly weighting
the returns within a time-window of length $T_1$ (different estimates
for different values for $T_1$):
\be
C_{ij}^{h,eq,T_1}=\frac{1}{T_1}\sum_{k=0}^{T_1-1} x_{ik}x_{jk},
\ee
where $x_{ik}$ denotes the return on asset $i$ at time $t-k$,
with $t$ being the last point of the estimation period.
We also calculate "historical" estimates
based on exponentially weighting of returns (different estimates 
for different values for the decay-factor $\alpha$):
\be
C_{ij}^{h,exp,\alpha}= (1-\alpha)
\sum_{k=0}^{T_s-1} \alpha^k x_{ik}x_{jk},
\ee
where $T_s$ denotes the length of the estimation period 
(which is chosen so that $\alpha^{T_s}\ll 1$).
In addition, we consider covariance estimators based on "filtering"
the historical (uniformly and exponentially weigthed) matrices. 
For each historical matrix we consider two versions of filtering: one,
based on the largest eigenvalue\footnote{Such procedure
is consistent with the "single index" or "market" model widely
used by practitioners.}, and the other, based on the eigenvalues above the 
noise band of the corresponding random matrix.
Let us denote these by
$C_{ij}^{m,eq,T_1}$, $C_{ij}^{m,exp,\alpha}$ and by 
$C_{ij}^{r,eq,T_1}$, $C_{ij}^{r,exp,\alpha}$, respectively.
To summarize, for each value of $T_1$ we have three estimators based on
uniformly weighting the returns, and for each value of $\alpha$ we have
three estimators based on exponential weighting. 

In what follows, we compare the performance of these estimators for 
different values of $N$ (number of stocks) and $T_2$ (length of
"investment period"). The criteria used for comparison is the
ex-post volatility (i.e.\ the volatility in the "investment period") 
of the minimum variance portfolios constructed by using the estimators
based on ex ante return data (i.e.\ before the investment period).
The volatility measures are obtained by averaging over a large number
of bootstrapped samples obtained from the dataset of daily stock returns.
The results for $N=100$ and $T_2=20$ (investment period of one month)
are presented in Fig.\ \ref{fig:n100ttwo20}.

\begin{figure}
\begin{center}
\includegraphics[scale=0.65]{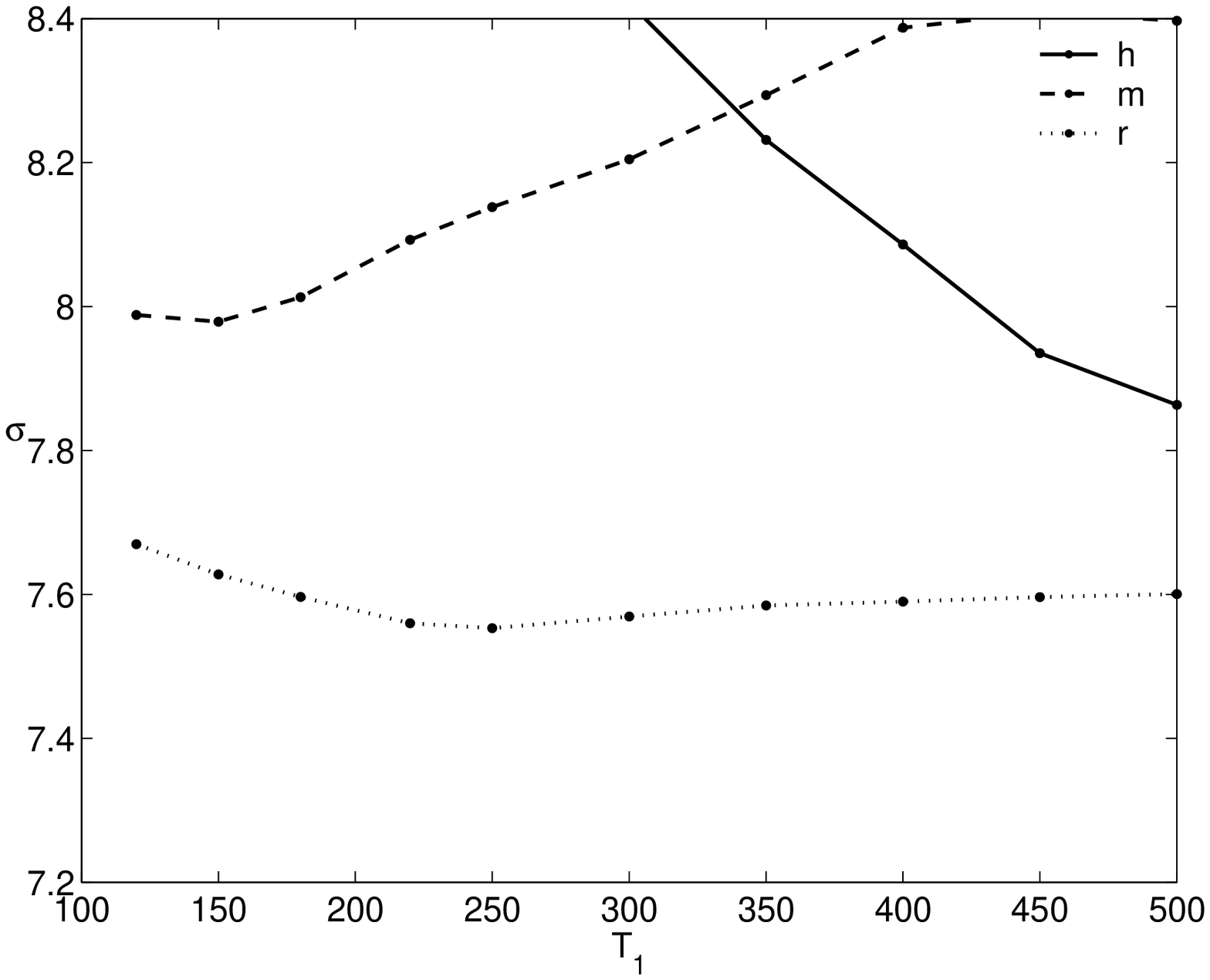}
\includegraphics[scale=0.65]{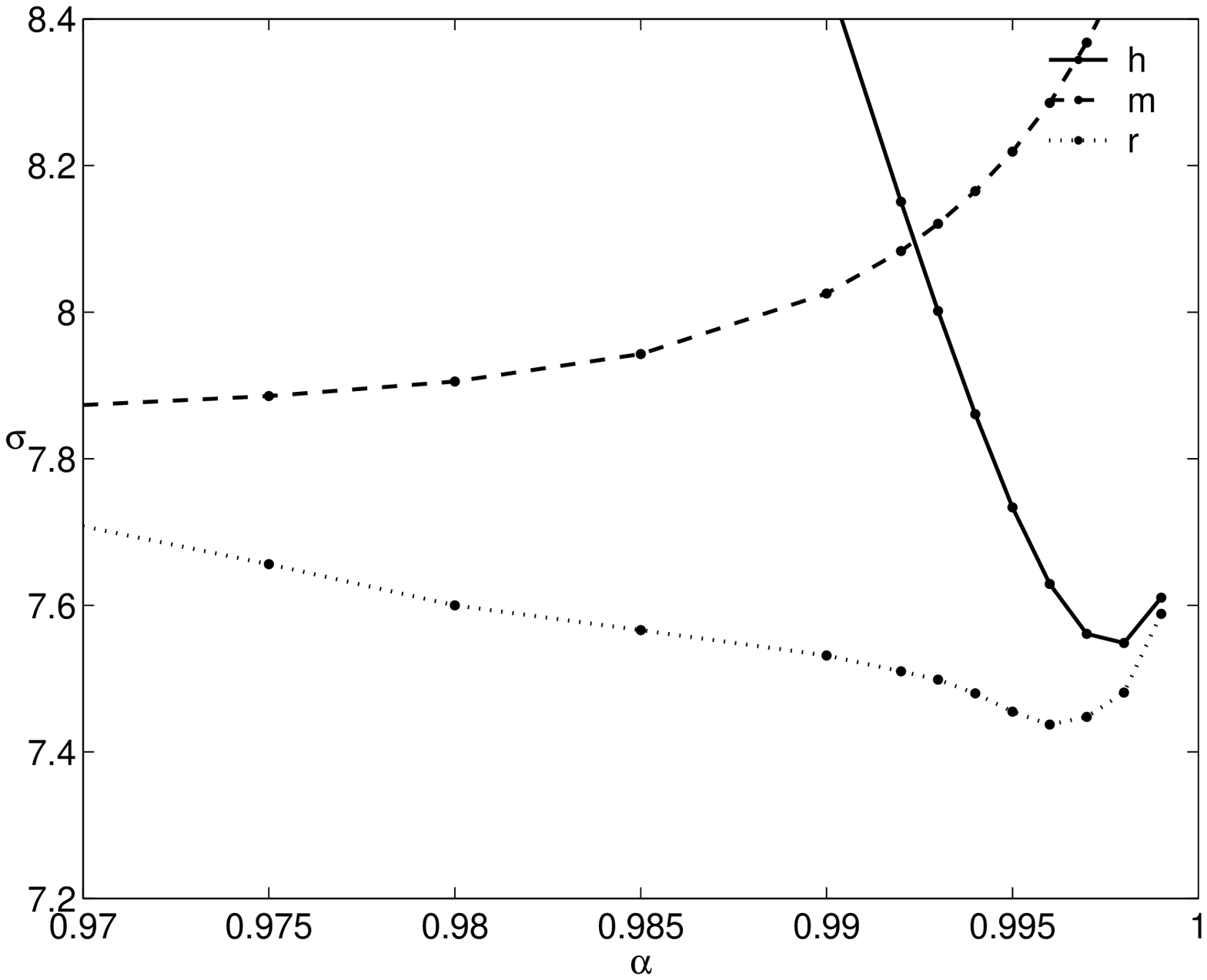}
\end{center}
\caption{Ex post volatility (annual \%) of optimal portfolios
constructed using correlation matrix estimators based on
{\it (top)} uniform weighting, {\it (bottom)} exponentially weighting 
the return data, as a function of 
{\it (top)} the length $T_1$ of the time window, {\it (bottom)} the decay factor
$\alpha$ used in the weighting. {\it (h)}, {\it (m)}, {\it (r)}
denote the results obtained in the case of {\it (h)} historical/sample
estimate, {\it (m)} single-index/market model estimate and
{\it (r)} estimate using random matrix theory-based filtering, respectively.
\label{fig:n100ttwo20}}
\end{figure}

It can be seen from the figure that the {\sl random-matrix-theory-based
filtering} performs the best for both uniform and exponential
weighting. It is interesting to note that in the case of uniform weighting 
the best choice for the length of the time window $T_1$ is around 250,
i.e.\ one year of (daily) data. In the case of exponential weighting
the best choice of the parameter $\alpha$ is around 0.996, which 
corresponds to an effective time length of $-1/\log\alpha$ of around 250 again.
By comparing the two minima, one can see that the estimator based
on exponential weighting performs (slightly) better. This shows that
combining techniques that take into account the volatility and
correlation dynamics of time series (e.g.\ exponential weighting) 
with techniques that reduce the effective dimensionality of the 
correlation matrix (e.g.\ random matrix theory-based filtering) can provide 
covariance matrix estimates that lead to optimized portfolios with better 
risk characteristics.

The {\it historical estimators} can perform quite well if enough data points
are taken into account (i.e.\ $T_1$ or $\alpha$ is large enough).
For uniform weighting it seems that 2 years of daily data can be
enough (for $N=100$ assets!). One important point to note, however, is
that covariance matrices obtained using the RiskMetrics 
\cite{riskmetrics} method, i.e.\ exponentially weighted historical
estimate with $\alpha=0.94$, are completely inappropriate for
portfolio optimization with a larger number of assets. For example,
for $N=100$ the volatility of the optimal portfolio obtained by using
this matrix is around 16 (annu.\ \%), well above the values presented in
Fig.\ \ref{fig:n100ttwo20}. Even for $\alpha=0.96-0.97$ as used by
many practitioners, the volatility value is above 12 (annu.\ \%).
Therefore, although RiskMetrics has been found very useful in estimating
the market risk of portfolios, the results above suggest that its 
direct use for portfolio optimization with a larger number of assets may be
completely misleading. As a matter of fact, this seems to have been realized by
practitioners, who advocate e.g.\ the use of larger $\alpha$ \cite{goldman} or
of principal component analysis \cite{factorarch} (which, in view of our
results, can be interpreted as increasing the effective time length or 
decreasing the dimensionality of the problem, respectively).

It is interesting to note that {\it single-index estimators} 
perform better when a smaller number of data points is used for the estimation.
The reason for this could be that the fewer data points are used, the more
correlation dynamics can be taken into account, while the loss
of the estimation precision for the largest eigenspace is probably smaller.

Simulations for longer "investment periods" (larger $T_2$) showed
very similar results. For more assets (larger $N$) results are similar,
although the effectiveness of historical estimates decreases further.
However, for fewer assets (e.g.\ $N=50$) historical estimates perform
better and can compete with estimates based on random matrix theory 
filtering.


\section{Conclusion}

We have introduced a covariance matrix estimator that takes into account the heteroskedastic
nature of return series and reduces the effective dimension of portfolios (hence measurement noise) via
techniques borrowed from random matrix theory. We have demonstrated its superiority to both the method 
of exponentially weighted moving averages and the uniformly-weighted random-matrix-theory-based filtering.
We have found that a too strong exponential cutoff will waste too many data, while a weak cutoff will
wash away the non-stationary nature of the time series. The optimal attenuation factor, corresponding 
to the best balance between these two extremes, was found to be higher than
the value suggested by RiskMetrics.


\section*{Acknowledgements}

This work has been supported by the Hungarian National Science
Found OTKA, Grant No.\ T 034835. We are grateful to J.-P.\
Bouchaud and K. Giannopoulos for valuable discussions.



\end{document}